\documentclass[preprint,12pt,numbers,sort&compress]{elsarticle}

\usepackage{amsmath}
\usepackage{amsfonts}
\usepackage{amssymb}
\usepackage{txfonts}
\usepackage{natbib}
\usepackage{siunitx}
\usepackage{lineno}
\usepackage{subcaption}
\usepackage{graphicx}
\usepackage[colorlinks,
            linkcolor=blue,
            citecolor=blue,
            filecolor=blue,
            urlcolor=blue]{hyperref}
\usepackage[export]{adjustbox}
\usepackage{soul}

\journal{Astroparticle Physics}

\begin{document}

\begin{frontmatter}


\title{Atmospheric Cherenkov Telescopes as a Potential Veto Array for Neutrino Astronomy}

\author[1]{D.~Rysewyk\corref{cor1}\fnref{fn1}}
\cortext[cor1]{Corresponding author}
\fntext[fn1]{Biomedical and Physical Sciences Building, Michigan State University, 567 Wilson Rd. Room 4226, East Lansing, MI 48824, USA}
\ead{rysewykd@msu.edu}
\author[1]{D.~Lennarz}
\ead{dirk.lennarz@pa.msu.edu}
\author[1]{T.~DeYoung}
\ead{tdeyoung@msu.edu}
\author[2]{J.~Auffenberg}
\ead{auffenberg@physik.rwth-aachen.de}
\author[2]{M.~Schaufel}
\ead{merlin.schaufel@rwth-aachen.de }
\author[2]{T.~Bretz}
\ead{thomas.bretz@physik.rwth-aachen.de}
\author[2]{C.~Wiebusch}
\ead{wiebusch@physik.rwth-aachen.de}
\author[1]{M.~U.~Nisa}
\ead{nisamehr@msu.edu}

\address[1]{Department of Physics and Astronomy, Michigan State University, East Lansing, MI, USA}
\address[2]{III. Physikalisches Institut, RWTH Aachen University, D-52056 Aachen, Germany}

\begin{abstract}
The IceCube Neutrino Observatory has revealed the existence of sources of high-energy astrophysical neutrinos. However, identification of the sources is challenging because astrophysical neutrinos are difficult to separate from the background of atmospheric neutrinos produced in cosmic-ray-induced particle cascades in the atmosphere. The efficient detection of air showers in coincidence with detected neutrinos can greatly reduce those backgrounds and increase the sensitivity of neutrino telescopes.  Imaging Air Cherenkov Telescopes (IACTs) are sensitive to gamma-ray-induced (and cosmic-ray-induced) air showers in the 50~GeV to 50~TeV range, and can therefore be used as background-identifiers for neutrino observatories. This paper describes the feasibility of an array of small scale, wide field-of-view, cost-effective IACTs as an air shower veto for neutrino astronomy. A surface array of 250 to 750 telescopes would significantly improve the performance of a cubic kilometer-scale detector like IceCube, at a cost of a few percent of the original investment. The number of telescopes in the array can be optimized based on astronomical and geometrical considerations.
\end{abstract}


\end{frontmatter}


\section{Introduction}

A flux of high-energy neutrinos from astrophysical sources was discovered by the IceCube Neutrino Observatory in 2013 \citep{bib:HESE}.  Although evidence has been presented for several possible multi-messenger correlations  \citep{Kadler:2016ygj,Padovani:2016wwn,Lucarelli:2018jwo,Aartsen:2019gxs}, the only source identified with high confidence to date is the blazar TXS 0506+056 \citep{IceCube:2018dnn,bib:TXSBlazar}.  However, IceCube limits on the total contribution of GeV blazars to the observed astrophysical neutrino flux \citep{bib:blazar_stacking} imply that the majority of the neutrino flux is produced in other, as-yet-unidentified class(es) of sources.  

If blazars are not the primary sources of the neutrino flux, it is challenging to reconcile IceCube measurements of the neutrino flux below 200~TeV with the diffuse extragalactic gamma-ray background measured by Fermi at GeV scales \citep{Ackermann:2014usa}.  Most models of astrophysical neutrino sources predict that the neutrinos are produced in the decay of $\pi^\pm$ and $K^\pm$ mesons.  Neutral mesons will be produced alongside the charged mesons, and decay to produce gamma rays.  Although high-energy gamma rays can be reprocessed through interaction with the cosmic backgrounds of infrared and microwave photons, the integrated energy emitted in high-energy photons and neutrinos is tightly coupled.  The soft spectral index observed with IceCube at energies below 200~TeV \citep{Aartsen:2015knd,Aartsen:2016xlq} implies a higher bolometric neutrino luminosity than can be accommodated easily in the non-blazar component of the Fermi diffuse gamma-ray background \citep{TheFermi-LAT:2015ykq}.  Options for resolving this tension include a sharp break in the neutrino spectrum at energies just below those currently accessible to IceCube; the existence of a significant but currently unidentified Galactic component in the IceCube flux; neutrino production in some currently unknown class of objects with opacity large enough to process the photons to energies below the gamma-ray band (or via new physics such as dark matter); or considerable errors in the fraction of either neutrinos or gamma rays attributed to blazars \citep{bib:murase,Murase:2015gea,Murase:2015xka,bib:chang_wang,bib:tamborra,bib:ando,bib:chang,Kopp:2015bfa,Wang:2015mmh,Hooper:2016jls,Dev:2016qbd,bib:energy_budget_1,Linden:2016fdd,Bhattacharya:2016tma,Denton:2017csz,Palladino:2018evm,Sui:2018bbh}.  Speculation regarding the possibility of features in the neutrino energy spectrum in the 10--100 TeV range \citep{Bhattacharya:2014vwa,Bhattacharya:2017jaw,Chianese:2016kpu,Chianese:2017nwe}, and the possibility of using the inelasticity of events observed at a few tens of TeV to measure the neutrino/antineutrino ratio and thus determine whether astrophysical neutrinos are generated in hadronuclear ($pp$) or photonuclear ($p\gamma$) processes in their sources \citep{Anchordoqui:2004eb,Hummer:2010ai,Bhattacharya:2011qu, Xing:2011zm,Barger:2014iua,Shoemaker:2015qul,Biehl:2016psj,Nunokawa:2016pop,Aartsen:2018vez}, further contribute to the interest in improved measurements in this energy range.

Improved methods for identifying astrophysical neutrinos and reducing backgrounds at energies below the 100~TeV scale are thus of considerable interest.  Neutrino telescopes such as IceCube \citep{bib:IceCube}, ANTARES \citep{bib:ANTARES}, the Baikal Neutrino Observatory \citep{bib:Baikal} as well as the planned Baikal-GVD \citep{bib:GVD} and KM3NeT \citep{bib:KM3NeT} use photomultiplier tubes (PMTs) to detect Cherenkov radiation from charged leptons produced through neutrino interactions in the surrounding ice or water. 
Charged-current (CC) muon neutrino interactions produce track-like events in a neutrino telescope. Charged-current electron and tau neutrino interactions, as well as neutral-current (NC) neutrino interactions of any flavor, produce cascade-like events. Cherenov light emission from a cascade is nearly spherically symmetric, while tracks are elongated due to the muon path length, which is typically several kilometers for TeV-scale muons in ice or water.

The main backgrounds in searches for extraterrestrial neutrinos are ``atmospheric'' muons and neutrinos, which are produced in cosmic-ray-induced particle cascades in the atmosphere known as air showers.  The classic technique to reject atmospheric muons is to search for particles coming upward through the Earth from the opposite hemisphere, so that only neutrinos can reach the detector. However, this method cannot distinguish atmospheric neutrinos from those produced in astrophysical sources, except on a probabilistic basis based on the typically softer energy spectrum of atmospheric neutrinos.  

A second technique for rejecting atmospheric backgrounds is to use entering atmospheric muons detected in the outer region of the neutrino telescope as indicators of air showers, enabling both muons and downward-going atmospheric neutrinos to be vetoed \citep{bib:DOM_veto,Gaisser:2014bja,Arguelles:2018awr}.  This is the method exploited by the IceCube Collaboration to first discover the diffuse astrophysical neutrino flux \citep{bib:HESE}. A variation of this method is to detect the air shower itself with a separate array of detectors at the surface.  In IceCube, this approach has been demonstrated using the IceTop array of (frozen) water Cherenkov particle detectors on the surface above IceCube \citep{bib:IceTopVeto}, but only above a cosmic-ray energy threshold\footnote{The energy of an atmospheric neutrino is a fraction of the energy of the primary cosmic ray.  Although neutrino production peaks later in the shower development, the highest-energy neutrinos arise from mesons produced in the first generations of the air shower.  The steeply falling cosmic-ray spectrum implies that for a given {neutrino} energy $E_\nu$, cosmic rays with energies $E_p$ only a few times higher than $E_\nu$ contribute significantly.  As a rough rule of thumb, for an array with a cosmic-ray energy threshold of $E_p$, atmospheric neutrinos can be vetoed efficiently above an energy $E_\nu \gtrsim E_p /3$.}  of 1~PeV and over a relatively small field of view.  

In this paper, a potential array of Imaging Air Cherenkov Telescopes (IACTs) as a dedicated surface air shower detector array for IceCube is evaluated. The ability for IACTs to perform as a veto array with an efficiency and energy threshold that is relevant for neutrino astronomy is assessed. IACTs detect air showers by measuring Cherenkov light emitted from the particles during an extended air shower. Although the requirement of dark, clear skies (to reduce background and boost signal) will limit the duty cycle of such an array, IACTs offer the potential advantages of a lower energy threshold and larger ground coverage per station compared to direct particle detection with surface arrays such as IceTop.  To be relevant for neutrino astronomy, the array must be capable of detecting high-energy cosmic-ray air showers with very high efficiency over a surface area of order 1 km$^2$ and a solid angle of at least a large fraction of a steradian.  The ability to operate, even with reduced sensitivity, in harsh environments with bright Moon or aurora conditions is important to maximize the potential duty cycle.  

The IACTs required to veto atmospheric neutrinos need not be as sophisticated as those used for gamma-ray astronomy, however.  An IACT of $\sim$ 0.5~m$^2$ collection area is easily capable of detecting air showers at or below the 100~TeV scale, which produce essentially all atmospheric neutrinos with energies above a few tens of TeV \citep{bib:HAWCeye}.  Angular resolution is not crucial: the Cherenkov radiation from air showers is strongly beamed, so only showers aimed generally at the IACT and the neutrino telescope below it will be detectable, and the potential transverse momentum of atmospheric neutrinos limits the utility of angular resolution better than a few degrees.  Discrimination of gamma-induced and hadron-induced showers, a key performance metric for gamma-ray astronomy, is irrelevant in this context.  

For this study, we consider an array of ground stations, each consisting of multiple small IACTs in a ``fly's-eye'' arrangement.  We base performance assumptions on a design for an enclosed telescope with a camera of silicon photomultipliers (SiPMs), suitable for harsh environments and capable of operation in relatively high ambient light levels \citep{bib:Bretz}.  Prototypes of similar design have been operated at the South Pole in coincidence with IceCube \citep{bib:iceact_demonstrator}.  

This paper is structured as follows. Section~\ref{sec:telescope} presents the conceptual design for the telescopes which make up the array.  Section~\ref{sec:simulations} describes air shower simulations used to characterize the performance of the telescope as an air shower detector. Section~\ref{sec:array} outlines a possible veto array which could be installed at the IceCube Observatory. Section~\ref{sec:neutrinoastronomy} discusses the potential impact of such an array on neutrino astronomy.

\section{Conceptual Design of a Small-Scale IACT}\label{sec:telescope}
\subsection{Optics and Photon Collection Efficiency}

\begin{figure}[tp!]
    \centering
    \includegraphics[width=0.65\linewidth]{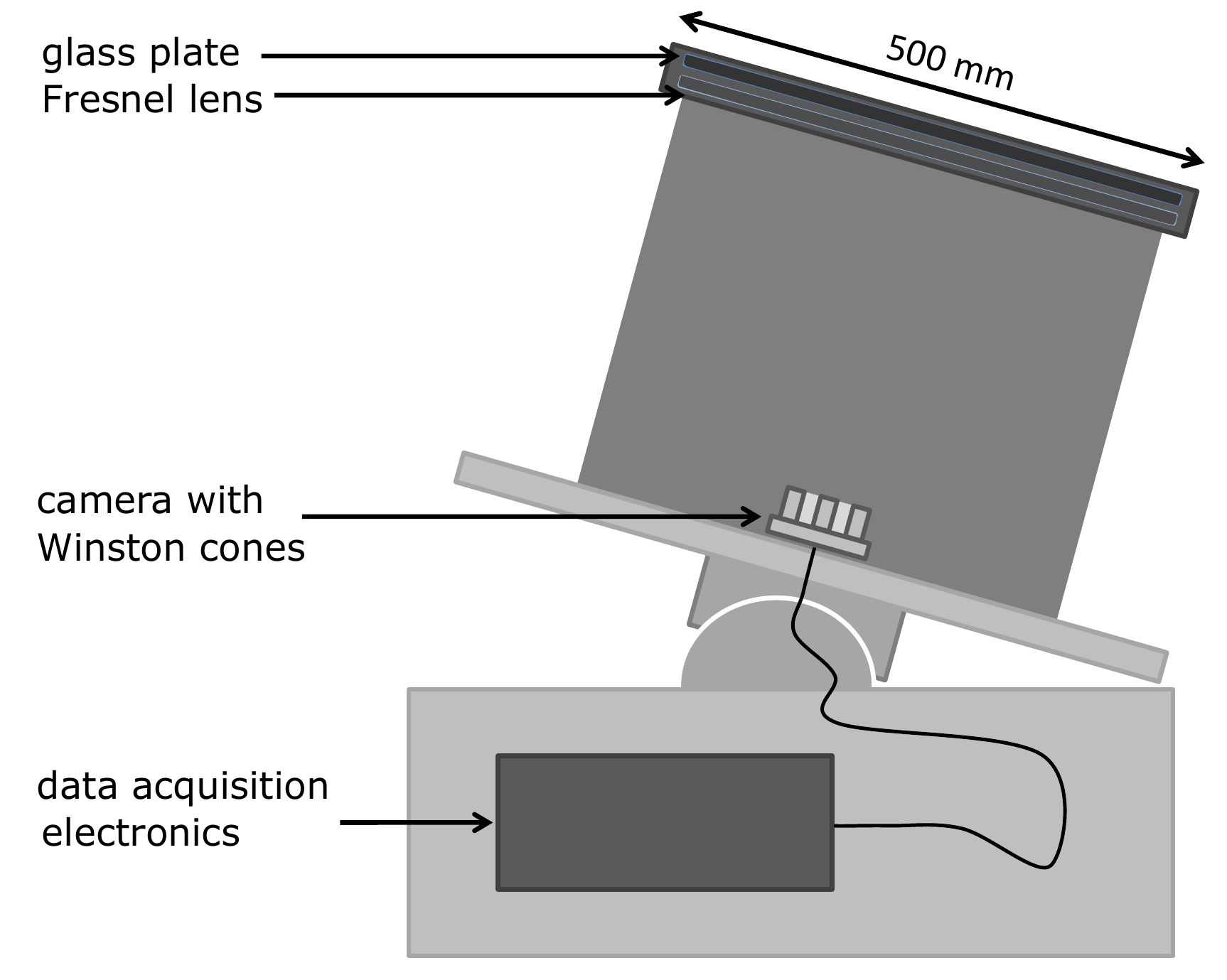}
    \caption{An example of a small-scale IACT adapted from \cite{bib:iceact_icrc}. It features a Fresnel lens and a camera based on 61 SiPMs with Winston cones.}
  \label{fig:iceactdiagram}
\end{figure}

This study envisions an array of IACTs similar to the IceACT telescope \citep{bib:Bretz,bib:iceact_icrc,bib:iceact_demonstrator}, which has a light collection area of $0.237~\rm{m}^2$ and a field of view of 0.045~sr (roughly $14^\circ$ diameter). An example diagram of one of these IACTs can be seen in Figure~\ref{fig:iceactdiagram}. It features a large Fresnel lens as imaging optics with a diameter of $D=500$~mm and a numerical aperture, defined as the ratio of the focal length $F$ to the diameter, of $F/D\sim 1$. The camera is based on 61 SiPMs, which have been widely explored as an alternative to classical PMTs in IACTs because they offer gain and quantum efficiency comparable to PMTs but can be operated under much brighter light conditions \citep{bib:FACT,bib:biland}. Winston cones are used in the focal plane to increase the effective photon detection area \citep{bib:janpaul,bib:maurice}. The total cost of such a telescope is estimated to be less than 10,000 euros \citep{bib:Bretz}.

Since IACTs can only operate in low light conditions, the telescopes can operate during astronomical night which corresponds to winter in the Southern hemisphere.  At South Pole, these conditions obtain for 4 months of the year, while the Sun is more than 12.6$^\circ$ below the horizon\footnote{Due to the lower aerosol content of the atmosphere over the Antarctic plateau, sky darkening occurs more rapidly than at most mid-latitude sites.  Following Sims et al., we define astronomical night by the solar zenith distance at which the median moonless dark sky darkens by 0.5 mag.} \citep{bib:sims}.  Dark skies are required to detect air showers, although the SiPM-based camera will be capable of operating with the Moon above the horizon \citep{bib:fact_icrc}.  While very little precipitation falls at the South Pole, windblown snow crystals can also impact operations. Current prototypes are testing the effectiveness of resistive heating cables to improve snow removal during the winter months. We estimate that a winter duty factor of at least 60\% (corresponding to 2.4 months of operation per year) will be possible, but this estimate must be confirmed based on the experience operating the IceACT prototype at the South Pole.

\begin{figure}[tp!]
    \centering
    \includegraphics[width=0.65\linewidth]{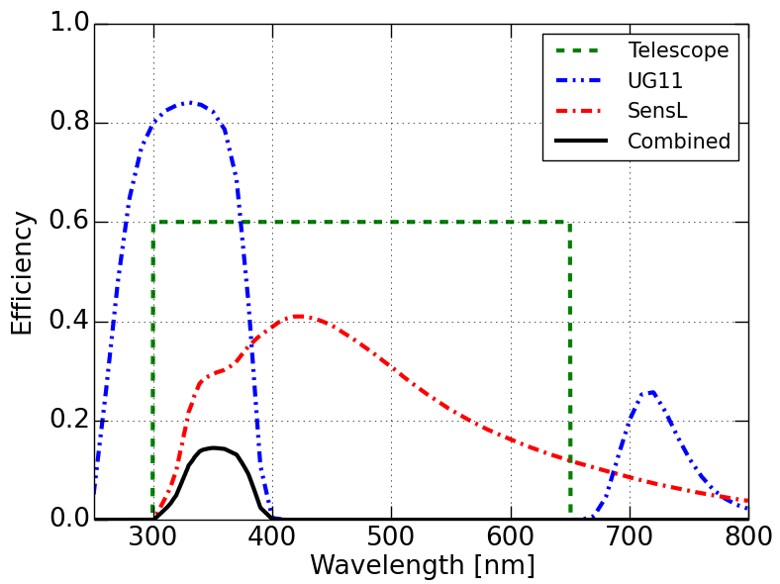}
    \caption{Performance of the various components of the telescope and their combined efficiency in the Cherenkov band. The dashed green line shows the photon collection efficiency assumed for the small-scale, wide field-of-view IACT. The dash-dotted red curve shows the photon detection efficiency (PDE) of the SensL SiPM, while the dash-double-dotted blue line shows the performance of the Schott UG11 filter. The black curve shows the overall efficiency of the telescope combining all of these components.}
  \label{fig:efficiency}
\end{figure}

We assume that the telescope optics provide a photon collection efficiency of 60\% between 300--650~nm.  This is an estimate of the efficiency that could be possible with a system based on a commercially-available Fresnel lens, but is not based on detailed modeling of a specific telescope design.
As a baseline for the camera we consider SensL-FJ ($\SI{6}{\mm}$\,x\,$\SI{6}{\mm}$) SiPMs (MicroFJ-60035-TSV) with an overvoltage of 5~V. For these parameters, the quantum efficiency is close to 30\% at 350~nm, as shown in Figure~\ref{fig:efficiency}. 

In addition to the usual diffuse night sky background (NSB), an IACT at the South Pole must also consider the aurora australis, discussed in detail in Section~\ref{sec:simulations}.  For this study we assume that a filter made of Schott UG11 glass, with a peak efficiency for photons of $\sim$90\% around 325~nm, is added to the telescope to attenuate the longer wavelengths emitted by the aurora while retaining the shorter wavelengths which predominate in the Cherenkov spectrum. 

The overall photon detection efficiency is the product of the  efficiency of the telescope optics, the UG11 filter efficiency, and the SensL SiPM efficiency.  The wavelength dependence of the efficiencies of various telescope elements, and the overall photon detection efficiency we assume in this study are given in Figure~\ref{fig:efficiency}.  

\subsection{Electronics and Dark Noise}

Temperature variations affect the gain of SiPMs at the level of about 10\% per 10 ${}^\circ$C.  
Ambient temperature is relatively stable over the Antarctic night, with typical variation of $\pm 10$ ${}^\circ$C during the winter months and maximum recorded deviations $\pm 25$ ${}^\circ$C.  Gain variations due to temperature of the SiPM can be corrected by adjustment of the applied voltage \citep{bib:biland}, with residual gain variations at the 1\% level.  For the purpose of air shower detection, gain variation at this level is negligible compared to variations in ambient light levels.  

Dark noise in the SiPMs does not pose a major problem at the temperatures typical of the South Pole winter.  The single photon rate for SensL-J SiPMs is about $\SI{50}{\kHz/\mm}^2$ at 20 ${}^\circ$C, and falls by a factor of 3 for every 10 ${}^\circ$C. Typical operating temperatures at the South Pole in winter are around -50 ${}^\circ$C. The highest temperature recorded at the South Pole in winter is approximately -30 ${}^\circ$C, corresponding to a noise rate of $\SI{0.2}{\kHz/\mm}^2$.  For a 61-pixel camera with $\SI{6}{\mm}\times \SI{6}{\mm}$ sensors this corresponds to an overall dark count rate less than 
$\SI{450}{\kHz}$, or $\lesssim 10^{-2}$ per 10 ns (the time scale appropriate for Cherenkov light from air showers.  Cross talk in the SiPMs can produce apparent 2--3 photon signals, but at rates 1--2 orders of magnitude lower than the single count rate.  As these rates are more than two orders of magnitudes lower than the NSB levels discussed in Section~\ref{sec:simulations}, we do not consider dark noise further in this study. 
Bright stars in the field of view may also increase noise rates.  We assume that the impact will be limited to individual pixels and that realistic trigger electronics would be able to exclude high-rate pixels temporarily as a star drifts across the field.  As a detailed model of the trigger is beyond the scope of this study, we do not attempt to account for this effect.

To veto atmospheric neutrinos effectively, the telescopes in the array should be operated with the lowest possible trigger threshold.  For this study, we do not attempt to design a complete data acquisition system, but only to model the performance of the telescopes at a conceptual design level.  We assume the data acquisition electronics will be capable of recording single photoelectrons and implementing a simple coincidence logic which sums the total charge in the brightest three neighboring pixels of the telescope (the approximate size of an air shower in the camera) for the time windows of interest.  Since information from the veto array is relevant only if the neutrino telescope below the array has detected a particle, we envision a system which relies primarily on the neutrino telescope trigger to define time windows of interest.  

We assume that low-level trigger primitives from the IACTs can be stored in a look-back buffer until the neutrino telescope trigger is formed, and that the IACT data from the relevant time window would be examined for an excess signal above the background noise consistent with the presence of an air shower.  While the rate of the trigger primitives may be very high, the actual trigger rate and data bandwidth would be throttled by the neutrino telescope trigger, which typically runs at a few kHz.  

\section{Detection of Air Showers}\label{sec:simulations}

To investigate the feasibility of an IACT array as an air shower veto, we simulated the response of our conceptual IACT design to air showers produced by cosmic-ray primaries of different species, at energies ranging from 10~TeV to 200~TeV and zenith angles from 0$^\circ$ to 60$^\circ$.  Based on the Cherenkov light production and estimated background levels, we estimate the maximum distance from the shower axis at which our IACT concept could detect each shower, if it occurred within the IACT's field of view in a time window of interest identified by the neutrino telescope. In this feasibility study we do not include a detailed model of the IACT electronics or optics, but rather estimate detectability based on simple signal-to-noise considerations.  In general, air showers may produce signals in multiple IACTs, particularly if the shower axis is approximately equidistant from multiple stations.  This will tend to enhance the prospects for detection relative to the estimates presented here, which are based on the signals recorded in single telescopes.  On the other hand, more detailed modeling of the backgrounds, trigger electronics and detector optics could reduce the estimated efficiency compared to the idealized signal-to-noise calculation used in this initial study.

\subsection{Cherenkov Light Production in Air Showers}
The cosmic ray simulation program CORSIKA \citep{bib:CORSIKA} version 7.4005 was used to simulate air showers, with QGSJet-01c \citep{bib:QGSJet} 
for high-energy interactions and FLUKA 2011.2c.1 \citep{bib:FLUKA_1,bib:FLUKA_2} for low-energy interactions. 
All Cherenkov photons between 300--600~nm were recorded over a 2~km by 2~km readout area on the ground around the shower core. Wavelength dependence of the Cherenkov emission angle was 
ignored, as it has been shown that this level of detail is normally not needed for Cherenkov telescopes \citep{bib:sim_telarray}. 
The simulation used CORSIKA's South Pole August atmosphere model
and an observation level of 2,834~m above sea level. Particles were tracked down to energy thresholds of 50~MeV for hadrons and muons and 0.3~MeV for electrons and gamma rays, well below the threshold for Cherenkov radiation at sea level ($\sim20$~MeV for electrons and $\sim4$~GeV for muons).

Vertical air showers were simulated for 300 proton, 20 oxygen and 50 iron primaries at 100~TeV.  In addition, thirty 100~TeV proton showers were simulated at zenith angles between 10$^\circ$--60$^\circ$ at 10$^\circ$ intervals to investigate the impact of zenith angle. Thirty vertical proton showers were simulated for primary energies 10~TeV, 20~TeV, 50~TeV and 200~TeV to assess energy dependence.

\begin{figure}[tp!]
    \centering
    \includegraphics[width=0.65\linewidth]{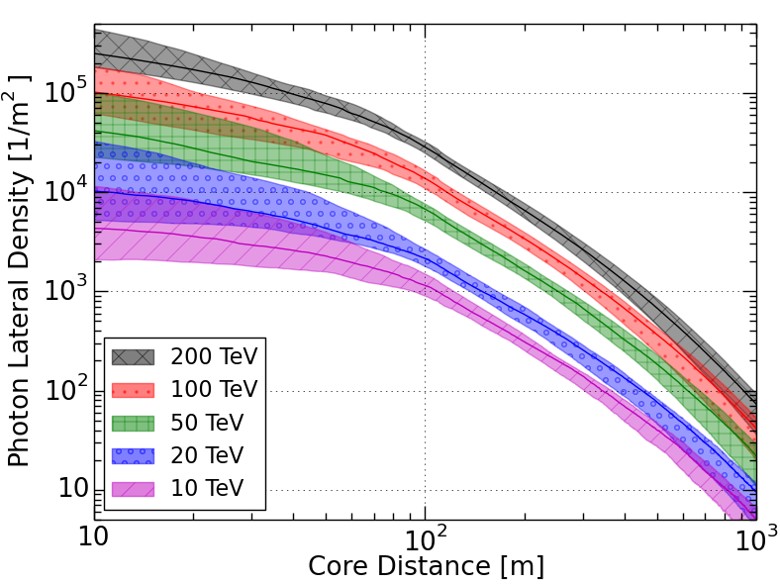}
    \caption{Mean lateral distribution of 300--600~nm Cherenkov photons for vertical proton primaries of different energies. A detector altitude of 2,834~m above sea level is assumed and effects of atmospheric absorption are not included. Each of the 30 showers is averaged in azimuth and the solid lines show the median of the averaged profiles. The shaded regions show the 15\% and 85\% percentiles for each energy.}
  \label{fig:light_distribution_energy}
\end{figure}

Figure~\ref{fig:light_distribution_energy} shows the mean lateral distribution of Cherenkov photons for vertical proton primaries of different energies. At energies below $\sim$1~TeV the photon density is known to be relatively flat and exhibits a characteristic rim at approximately 120~m \citep{bib:sim_telarray}. In the energy range of interest for the current study, the sharpness of this feature is greatly reduced, but a steepening in the rate of decline in photon density with distance can be seen. Typical Cherenkov photon densities for proton showers above 50~TeV are around 1,000~photons~m$^{-2}$~TeV$^{-1}$ close to the core, and 30~photons~m$^{-2}$~TeV$^{-1}$ at 200~m from the core.  Relatively large fluctuations from the median profile are common close to the shower core, but shower-to-shower variations are smaller at radial distances of 100--300~m.  

\begin{figure}[tp!]
    \centering
  \resizebox{0.65\hsize}{!}{\includegraphics{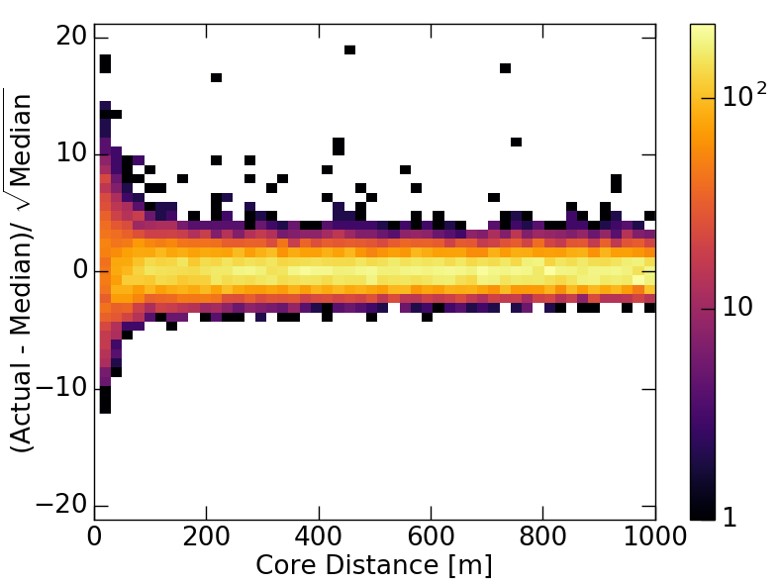}}
  \caption{Azimuthal fluctuations of the lateral distribution of 300--600~nm Cherenkov photons for a single vertical proton shower at 100~TeV. In 5~degree steps around the shower core and every 2~m in radius, a 0.5~m times 0.5~m square on the ground is selected and the actual number of photons is compared to the median profile that is obtained by averaging over all azimuth angles. The result is divided by the square root of the median, to determine whether the fluctuations are Poisson distributed (see discussion in main text).}
  \label{fig:fluctuations_sigma_paper}
\end{figure}

This effect is illustrated in Figure~\ref{fig:fluctuations_sigma_paper}, which shows the azimuthal variability within a single shower. For a Poisson process, the fluctuations should have a typical scale of $\sqrt{N}$, where $N$ is the expected number of photons.  Close to the shower core, where the number of Cherenkov photons is not well described by a Poisson process, photon densities are well above the threshold for detection so large downward fluctuations are unimportant.  At distances beyond about 100~m from the shower core, the photon density fluctuations are Poisson distributed, with the points in the plot having a mean close to 0 and a standard deviation close to 1. Small deviations caused by sub-structure in the air shower (e.g., energetic mesons with high transverse momenta $p_T$) create local fluctuations that are significantly higher than the expectation from the median profile, but downward fluctuations that could reduce detectability are well described by Poisson expectations.   
In this study we therefore assume Poisson distributions around the mean photon densities shown in Figure~\ref{fig:light_distribution_energy} for all Cherenkov photon counts.

\subsection{Atmospheric Extinction}
Some of the Cherenkov photons produced in air showers will be absorbed in the atmosphere  before reaching ground level.  The absorption probability depends on both the photon wavelength $\lambda$ and the heights of production and detection, and is tabulated in CORSIKA for a general atmospheric model. For vertical proton showers at 100~TeV, about 13\% of the Cherenkov photons between 300--600~nm are lost due to atmospheric extinction.  This number varies between about 8\% and 16\% according to the first interaction height of the primary particle: a shower that starts higher in the atmosphere will suffer more absorption before it reaches the ground.  The species of the primary particle also has some impact, since heavier primaries are less likely to penetrate deep into the atmosphere before the first interaction. For 30 vertical iron showers, the minimum and maximum atmospheric extinction are 15\% and 18\%.  Shower inclination also affects the absorption due to the increasing path length through the atmosphere, with extinction rising to 20\% for 100~TeV proton showers at 40$^\circ$ from zenith.

As discussed below in Section~\ref{sec:sensitivity}, the most challenging air showers to detect are vertical proton-induced showers which start deep in the atmosphere.  Nonetheless, we conservatively assume a value of 18\% extinction independent of energy and primary species. 
The resulting Cherenkov spectrum after atmospheric extinction is convolved with the total photon detection efficiency function shown in Figure~\ref{fig:efficiency} to determine the mean number of photoelectrons (PE) detected from each shower.  After all effects are taken into account, the mean PE yield near the shower axis is $\sim$550~PE~m$^{-2}$ of telescope area for vertical 100~TeV proton showers, or approximately 130 PE for our conceptual IACTs with 0.237~m$^2$ collection area.

\subsection{Sky Brightness and Aurorae}
\label{sec:aurora}

The Antarctic plateau is one of the world's premier sites for optical, infrared, and submillimeter astronomy, and NSB levels on the Antarctic plateau have received considerable attention from the astronomical community.  The plateau is a high-altitude desert with dark skies and relatively little cloud cover.  However, the aurora australis presents unique challenges which must be taken into account for an IACT veto array at the South Pole.  Aurorae are created when solar wind particles collide with the Earth's atmosphere, producing a diffuse glow in optical wavelengths which covers portions of the sky.  Fortunately, much of the auroral light is emitted at longer wavelengths which can be eliminated with a filter such as the Schott UG-11 glass included in our telescope model.  

Detailed characterizations of the aurora australis at both the South Pole and Dome A, roughly 1,000~km away on the peak of the plateau, have been reported in \cite{bib:sims} and \cite{bib:dempsey}.  Contributions to the NSB can be divided into two classes: line emission associated with the aurora australis, and continuum emission, including airglow, zodiacal light, starlight, diffuse galactic light, and quasi-continuous components of auroral emission ascribed to molecular bands of nitrogen and oxygen.  Spectral models of airglow and auroral emission and detailed discussions of the physical processes involved are presented in \cite{bib:gattingerII,bib:gattingerIII,bib:dempsey,bib:sims}.  

The intensity of the aurora vary considerably over the course of the year.  We used the median auroral activity presented in \cite{bib:sims} to calculate the impact of auroral activity on our telescopes.  Auroral intensity is usually described in terms of the intensity of the characteristic green emission from the 557.7~nm [OI] line, with the spectrum presented in  \cite{bib:sims} representing an IBC3 aurora event with a 557.7~nm line intensity of 100~kR.\footnote{1 Rayleigh (R) = $(10^{10}/4\pi)$ photons / $\rm{m}^2$ $\rm{s}$ $\rm{sr}$.} The six most dominant lines in the wavelength range of 300--650~nm are the 315.8, 337.0, 357.6, 391.4, 427.8, and 557.7~nm lines.  We consider all other auroral lines in this wavelength range to be subdominant.  We model the spectrum as scaling linearly with intensity, and normalize to the intensity of the 427.8~nm line:
\begin{align}
I_{315.8}&=0.40 \, I_{427.8}  \hspace{0.135in} I_{337.0}=0.82 \, I_{427.8}  \hspace{0.135in} I_{357.6}=0.55 \, I_{427.8}  \nonumber\\
I_{391.4}&=3.28 \, I_{427.8}  \hspace{0.135in} I_{557.7}=3.33 \, I_{427.8} \nonumber
\end{align}

Our model of the quasi-continuous component of the aurora is based on measurements from the Auroral Observatory in Fort Churchill, Manitoba \citep{bib:gattingerII,bib:gattingerIII}.  In the wavelength range of 310--470~nm, the continuum was found to have an intensity of 30~R/nm in reference to a 427.8~nm line intensity of 5~kR.  In the wavelength range of 450-890~nm, the continuum intensity was 270~R/nm in reference to a 557.7~nm line intensity of 100~kR.  For each of the benchmark auroral levels considered below, the continuum emission is dominated by these quasi-continuous auroral components.   We therefore scale the continuum background with the auroral intensity as well as the lines, following \cite{bib:dempsey}.

In order to quantify the effect of the varying auroral conditions on the performance of the IACTs, we define three benchmark aurora levels.  Observations of the brightness of the aurora australis were made at the South Pole Station during the winters of 1985 and 1990 \citep{bib:dempsey}, near solar minimum and maximum, respectively.  Auroral levels vary with the solar cycle, with solar minimum leading to higher auroral activity than during solar maximum.  The benchmark levels we consider are the mean auroral intensity during solar maximum, mean intensity during solar minimum, and the 90th percentile intensity during solar minimum to represent the highest intensities likely to be encountered on a regular basis.

The measured intensity of the 427.8~nm line at South Pole during both winters was reported in $B$~mags~$\rm{arcsec}^{-2}$, which can be converted to an intensity $I$ in photons~$\rm{m}^{-2}$~s$^{-1}$~sr$^{-1}$ using the equation \citep{bib:dempsey}
\begin{equation}
I = \frac{7.96\times 10^8 \;K \;\Delta \lambda \; 10^{(20-m)/2.5}}{1.24\times 10^6 \; \lambda},
\end{equation}
where $m$ is the brightness in magnitudes/$\rm{arcsec}^{2}$, $\lambda$ is the central wavelength of the filter in m, and $\Delta \lambda$ is the filter bandwidth in nm.  For the $B$ band, $K=42.6$, $\lambda=436\times10^{-9}$ m, and $\Delta \lambda=94$~nm. For the $V$ band, $K=36.4$, $\lambda=556\times10^{-9}$ m, and $\Delta \lambda=85$~nm.

\begin{figure}[tp!]
  \centering  
  \includegraphics[width=0.65\linewidth]{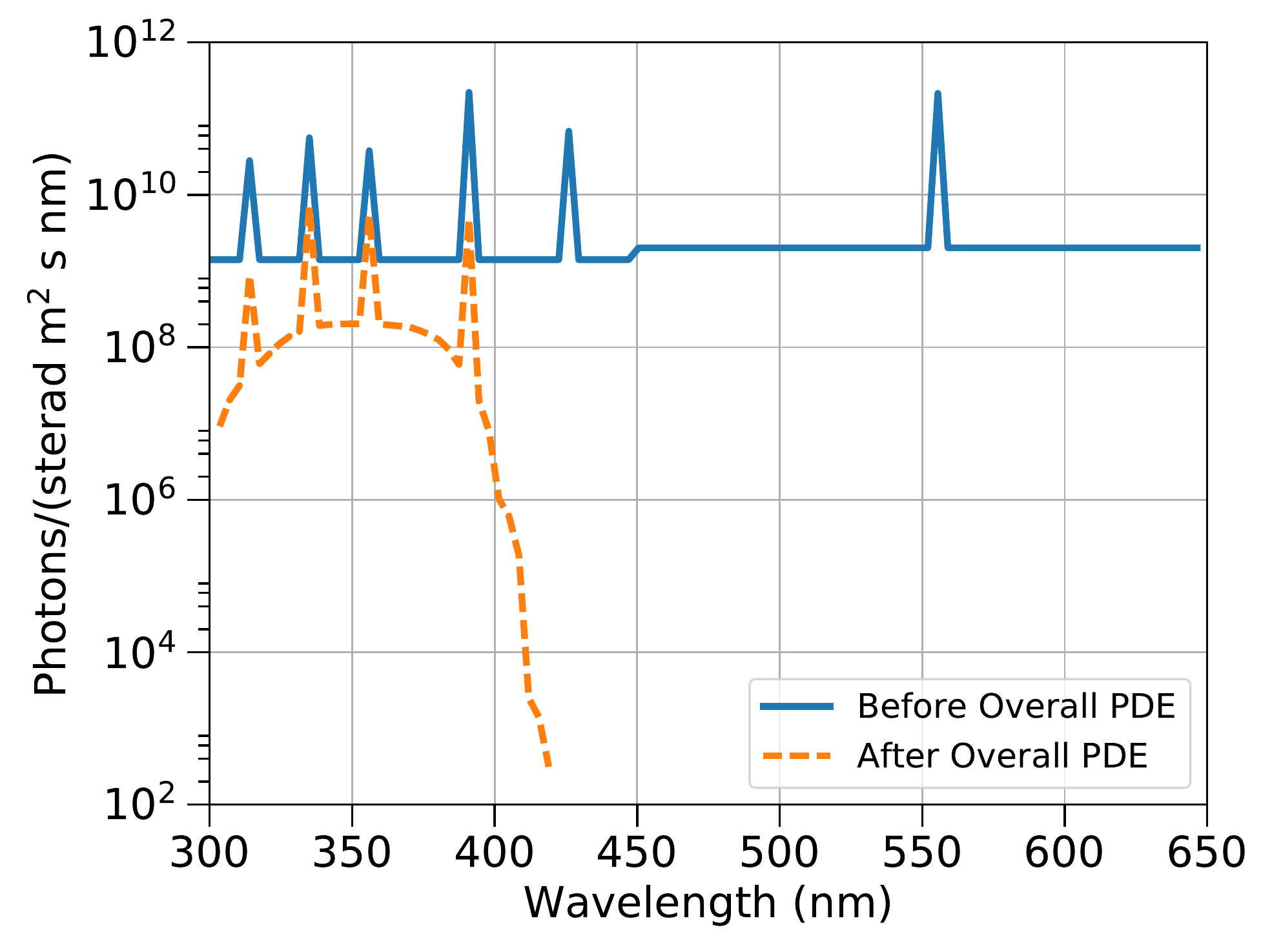}
  \caption{The spectrum of the NSB with mean-intensity aurora during solar maximum, before (solid blue line) and after (dashed orange line) applying the combined telescope photon detection efficiency from Figure~\ref{fig:efficiency}.}
  \label{fig:BeforeAfterPDE}
\end{figure}

The intensities of the 427.8 nm line at the three different benchmarks are 23.5 $B$ mags/$\rm{arcsec}^{2}$, 22.8 $B$ mags/$\rm{arcsec}^{2}$, and 21 $B$ mags/$\rm{arcsec}^{2}$ respectively.  The intensities of the other five major auroral lines and the continuum were scaled along with the 427.8 nm line intensity for each benchmark level, as explained above.  The background spectrum for the mean auroral intensity at solar maximum, including the continuum and six auroral lines, is shown in Figure~\ref{fig:BeforeAfterPDE} \footnote{Because the continuum has different intensities in different wavelength ranges, the different intensities were added together in a piecewise function at 450~nm based on \cite{bib:gattingerII} and \cite{bib:gattingerIII}. However, since the UG11 filter cuts out higher wavelengths, continuum emission levels in the wavelength range from 450-890~nm do not affect our estimates.}.

Applying the combined photon detection efficiency (PDE) shown in Figure~\ref{fig:efficiency}, the estimated sky background level at mean auroral intensity during solar maximum yields an average of 0.61 PE/ns detected across the full 61-pixel camera, for a $14^{\circ}$ FoV and 0.237 $\rm{m}^{2}$ collection area. This increases to 1.16 PE/ns for the background expected at the mean auroral intensity during solar minimum, and 6.1 PE/ns for the background from the 90th percentile intensity aurora during solar minimum.  Measurements from the IceACT prototypes are required to validate these estimates.


\subsection{Triggering}

Cherenkov photons from an air shower show dispersion in time of a few ns near the shower core. The distribution gets wider in time further from the shower core. For a given time distribution of photons, we identify the optimal start and stop times for a trigger window by maximizing the signal over the square root of the expected continuum background (during solar maximum).  This was repeated for 40 vertical proton showers. For distances of 140--160 m from the shower core, the optimal window was found to be 10 ns. The optimal time window retains 75\% of the original signal. For 190--210 m, the optimal window increases to 20 ns, with a similar fraction of signal retained.  


A complete design of the trigger electronics is beyond the scope of this study.  We assume that each telescope will form simple trigger primitives, and buffer the data from the corresponding time window until an external trigger decision is received from the neutrino telescope.  We model a simple coincidence logic that considers the total charge in the brightest three neighboring pixels of the telescope during the optimal time window identified above.  The charge threshold for the trigger primitives is determined by the need to limit accidental triggers due to background photons to a reasonable rate, which we define as less than 1 kHz.  

A toy Monte Carlo was used to calculate the coincident three-pixel PE distribution in a 20~ns time window for a telescope size of 0.237 $\rm{m}^{2}$, assuming that the background photons are uniformly distributed across the camera.\footnote{We anticipate that pixels with elevated rates, e.g. from a bright star transiting the field of view, would be temporarily excluded from the trigger.  Such effects are not included in this toy Monte Carlo.}  For a given trigger threshold, we computed the probability of accidental triggers based on the background levels calculated in Section~\ref{sec:aurora} for the three different auroral benchmark levels.  The probability of an accidental trigger in a 20~ns window with a threshold of $\geq$ 9 PE in three adjacent camera pixels for the lowest benchmark level (mean intensity during solar maximum) was found to be $5.3\times 10^{-6}$, corresponding to a rate of 265~Hz. The thresholds required to keep accidental trigger rates below 1~kHz for the mean and the 90th percentile intensity aurorae at solar minimum are found to be $\geq$11 PE and $\geq$24 PE respectively.

For air showers, the Cherenkov photon density at ground level scales approximately linearly with the shower energy, as shown in Figure~\ref{fig:light_distribution_energy}.  Comparison of these trigger thresholds thus implies that even very bright aurorae should shift the veto energy threshold only by a factor of two to three.  It should be noted that when bright aurorae are present, they often fill only a portion of the sky.  As high-energy atmospheric neutrinos are well aligned with their parent air showers and are reconstructed to degree-scale accuracy in IceCube, an IACT veto could remain effective over much of the sky even with bright aurorae obscuring air showers in some directions.  Experience with operating IACTs at the South Pole will be crucial for validating these expectations.  

\subsection{Air Shower Sensitivity} 
\label{sec:sensitivity}
From the Corsika simulation we obtain the number of Cherenkov photons per unit area as a function of distance from the shower axis. The photon counts are scaled down to account for atmospheric extinction before reaching ground level and the telescope optical efficiency.  We assume that 75\% of the photons will arrive within the trigger window and be collected within the three brightest camera pixels (each of which has a field of view of 1.5$^\circ$).  For each simulated shower, the lateral distribution of photons is used to calculate the maximum distance from the shower core at which the expected signal in the trigger window of a single IACT equals 9 PE, the required trigger threshold for typical operation during solar maximum.  This is the maximum distance from the telescope at which the simulated shower is considered detectable, which sets the maximum spacing between telescopes in a surface veto array.

\begin{figure}[tp!]
\centering
\begin{subfigure}{\linewidth}
    \centering  
    \includegraphics[width=0.65\textwidth]{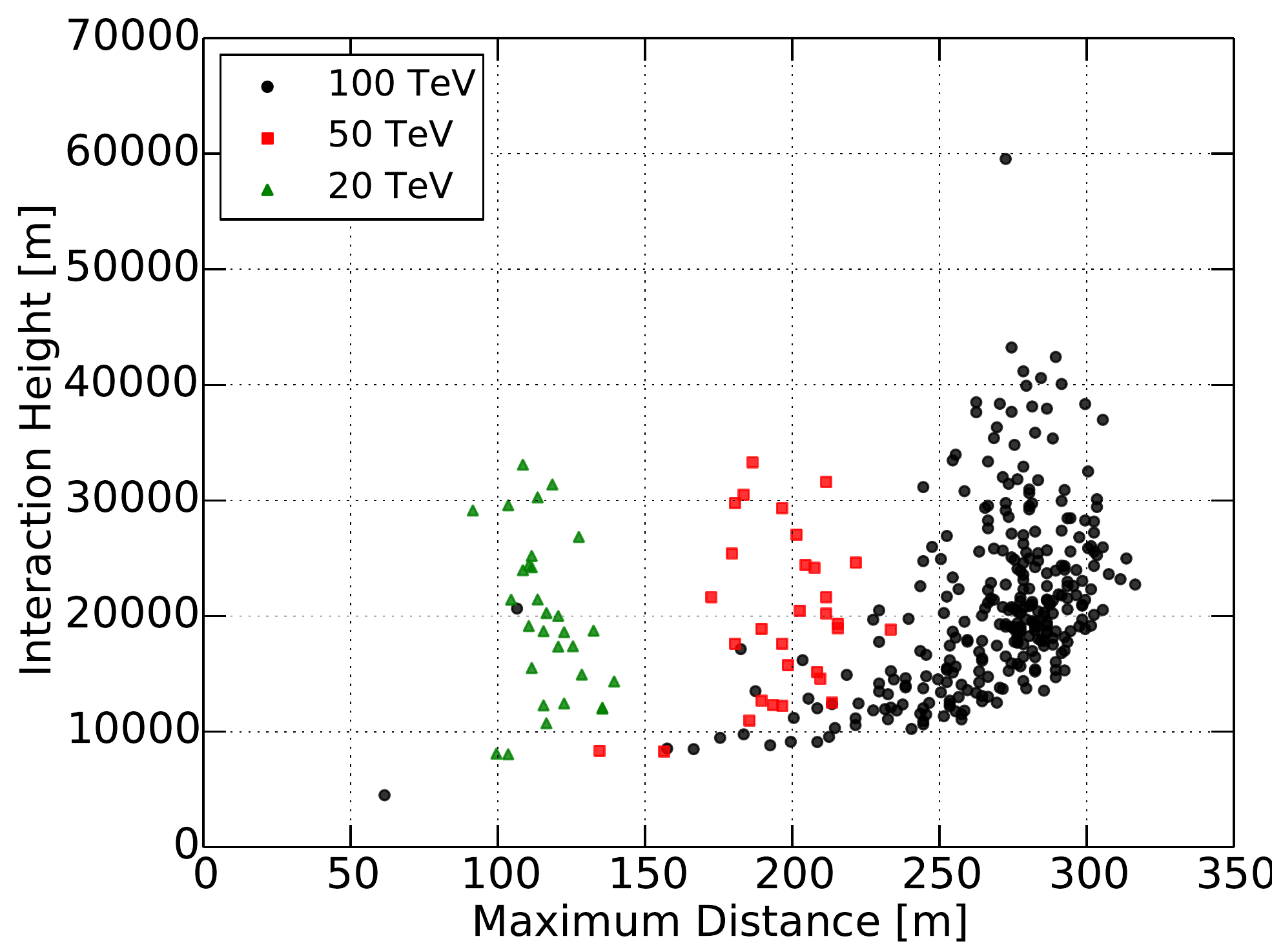}
    \caption{Maximum distance for detection vs.~first interaction height for 300 vertical proton-induced air showers at 100~TeV, compared to thirty showers each at 20~TeV and 50~TeV.}
    \label{fig:results_energy}
\end{subfigure}
\par\bigskip
\begin{subfigure}{\textwidth}
    \centering
    \includegraphics[width=0.65\textwidth]{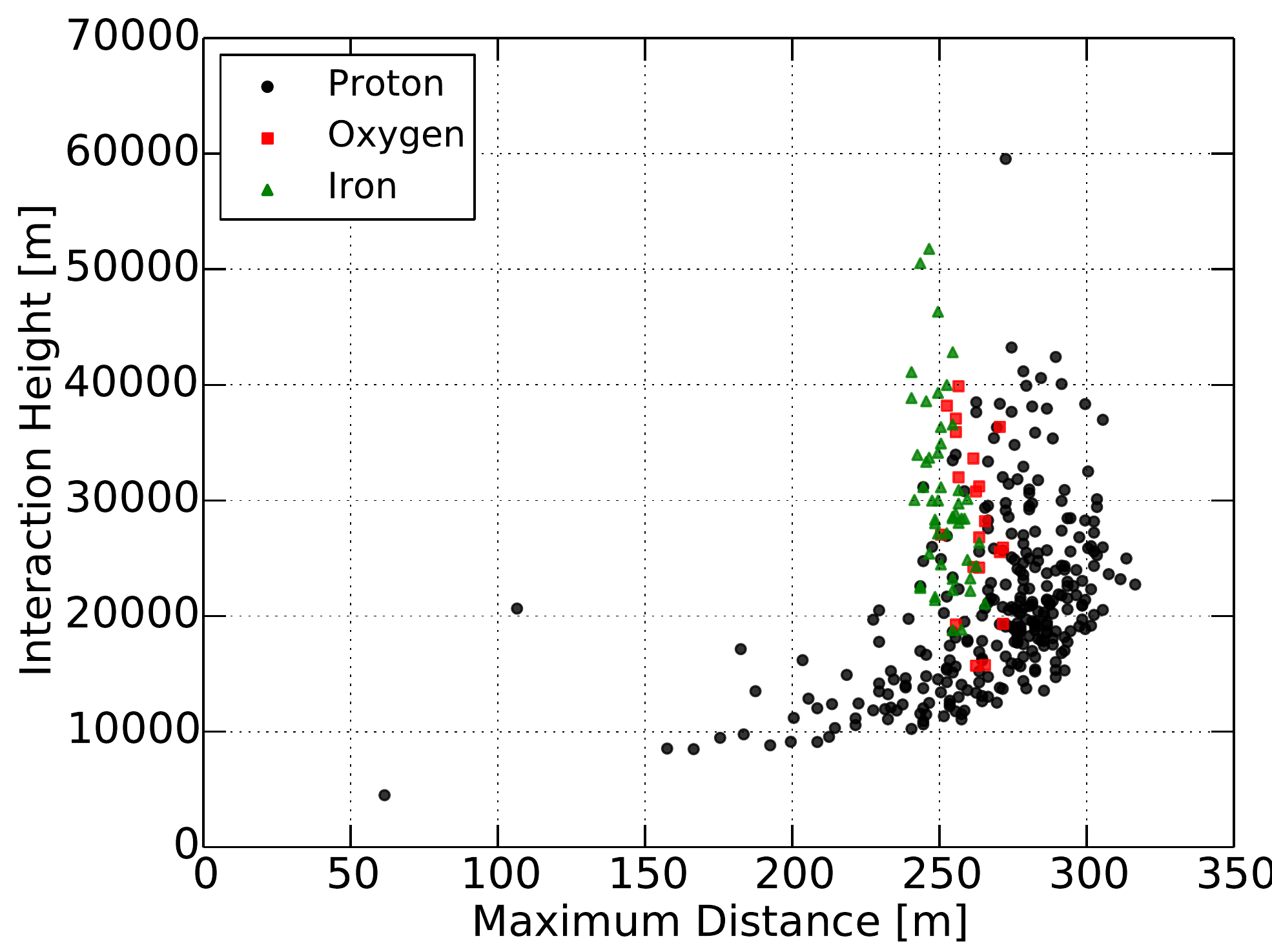}
    \caption{Maximum distance for detection vs.~first interaction height for different primary species.  Heavier nuclei are slightly dimmer but less likely to penetrate deeply into the atmosphere.}
    \label{fig:results_species}
\end{subfigure}
\caption{Maximum distance from telescope at which vertical air showers can be detected for a trigger threshold of 9 PE, plotted against the height of the first cosmic-ray interaction above the surface. Showers that penetrate deeply into the atmosphere are the hardest to detect.}
\label{fig:results}
\end{figure}

Figure~\ref{fig:results} plots the maximum detection distance for vertical showers of different energies and primary species against the height of the initial cosmic-ray interaction above the ground.  As seen in Figure~\ref{fig:results_energy}, the most important factor affecting detectability is the interaction height.  For cosmic rays which interact high in the atmosphere, the maximum distance at which showers of a given energy can be detected is generally consistent to roughly $\pm 25$\%.  The most difficult showers to detect are those in which the primary cosmic ray penetrates deeply into the atmosphere before interacting.  In such cases, the maximum detection distance is limited because there is not enough time for the air shower to develop fully and the Cherenkov light pool to expand laterally before reaching the ground.  Inclined showers are thus more reliably detected, as the increased slant depth through the atmosphere reduces the probability of very low interaction heights and the increased distance to the ground give more space for the Cherenkov pool to spread out.\footnote{Projection effects also improve the prospects for detecting inclined showers.  The telescope axis must be nearly parallel to that of the shower to observe the Cherenkov emission, so the density of photons in the telescope camera plane is essentially unaffected.  But the distance between telescopes in the plane perpendicular to the shower axis is smaller than in the ground plane, reducing the average projected distance from the shower axis to the nearest telescope.}  

For similar reasons, showers produced by heavier primaries are more reliably detected than those produced by protons.  As shown in Figure~\ref{fig:results_species}, the larger cross sections on air increase the typical height of first interaction, compared to proton primaries.  In addition, the air showers produced by heavy primaries develop more rapidly and more regularly than proton showers, so the Cherenkov light pool has more time to expand before reaching ground level.  Both effects cause the distance at which showers induced by heavy primaries of a given energy can be detected to be significantly more consistent than for protons.  

As shown in Figure~\ref{fig:results}, at primary energies of 100~TeV, $\sim$99\% of the vertical showers can be detected (and vetoed) up to 150~m away from the telescope.  Even at energies as low as $E_p = 50$~TeV, the detection/veto rate is $>$95\% for vertical proton showers.  Similar detection rates are also possible at 150~m from the telescope when the threshold is increased to $\geq$11~PE. However, if the threshold is increased to $\geq$24~PE, these detection rates are only achieved for showers up to 100~m from the telescope.  It should be noted that these efficiencies are for detection of the most challenging class of air showers: perfectly vertical proton-induced showers.  Detection rates for inclined showers and those produced by heavy nuclei will be higher than for this class of events.  We therefore conclude that our IACT concept should be able to detect air showers, and veto the associated atmospheric neutrinos, with at least 99\% efficiency to a radius of approximately 150~m above an threshold of 50--100 TeV.  This corresponds to an atmospheric neutrino energy threshold of approximately 15--30 TeV.  

We note that there may be correlations between atmospheric neutrino production in air showers and characteristics which enhance or suppress detectability by the IACT array at ground level.  For example, protons which penetrate too deeply into the atmosphere before first interaction may not leave enough time for energetic mesons to decay and produce neutrinos before reaching ground level, which would enhance prospects for detecting the showers producing high-energy atmospheric neutrinos.  Conversely, if a large fraction of the primary cosmic ray's energy is carried away by a neutrino, the electromagnetic component of the air shower may produce less Cherenkov light and be more difficult to detect.  More detailed air shower and telescope response simulations with higher statistics will be required to quantify these effects and further refine this estimate of the effective atmospheric neutrino energy threshold for veto efficiency.



\section{A Potential Air Shower Veto Array} \label{sec:array}

\begin{figure}[tp!]
  \includegraphics[width=0.35\linewidth]{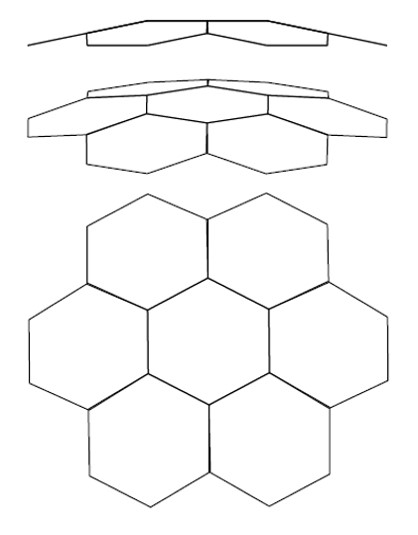}
  \centering
  \caption{Schematic arrangement of a telescope station (side view to top view). Each telescope has a hexagonal FoV of 14$^\circ$. Seven telescopes are arranged in a fly's eye array. The center telescope points straight up at 0$^\circ$ zenith. The six surrounding telescopes point at 14$^\circ$ zenith. With all seven telescopes, the station has a FoV of 36$^\circ$. }
  \label{fig:FlysEye}
\end{figure}

Based on these studies, we consider how one might implement an IACT-based surface veto array for the IceCube Neutrino Observatory.  To assist in the discovery of new neutrino sources, the veto array must cover a relatively large section of the sky.  This can be achieved by grouping multiple telescopes together into stations at different points above IceCube.  Each individual IACT has a $\sim$14$^\circ$ wide hexagonal FoV, and a station consists of up to seven telescopes arranged in a "fly's eye" configuration: one telescope pointing straight up, and six surrounding telescopes each pointing at a 14$^\circ$ angle from zenith.  With this arrangement, each station would have a FoV of up to 36$^\circ$ (i.e., extending to 17$^\circ$ from zenith). 
Figure~\ref{fig:FlysEye} illustrates the geometry of a station with seven telescopes.

\begin{figure}[t!]
    \centering
  \resizebox{0.9\hsize}{!}{\includegraphics{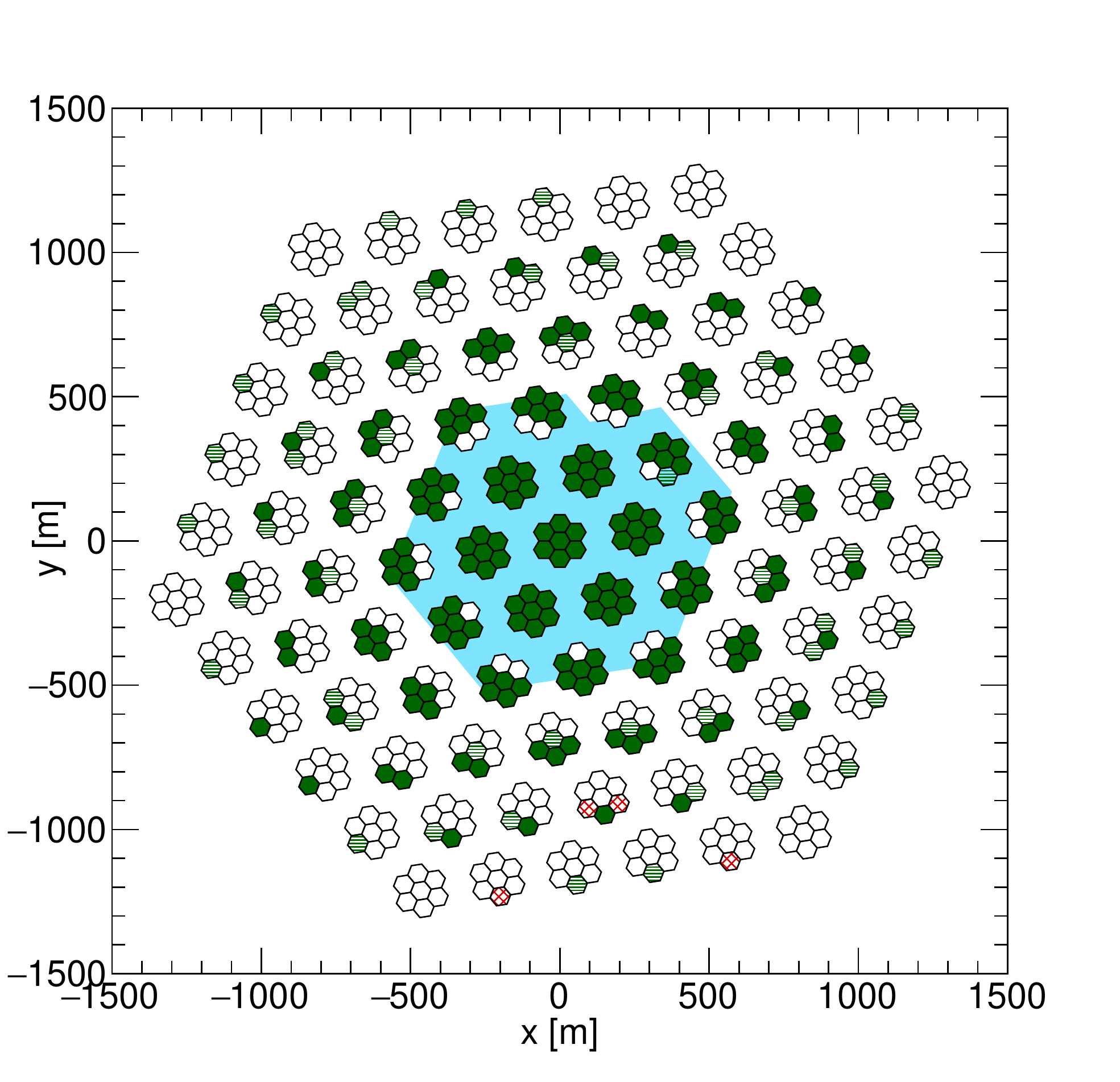}}
  \caption{Top view of a telescope array with $36^\circ$ FoV and a telescope station spacing of 260~m. The shaded blue region indicates the IceCube footprint. Each telescope in a station is colored according to how many DOMs are behind the telescope's FoV, from solid green ($>$50 DOMs) to green horizontal lines (11-50 DOMs), red crosshatched lines (1-10 DOMs) and white (none).}
  \label{fig:IceActLayout_fov18_white}
\end{figure}

A single station would have an effective coverage radius of 150~m in most light conditions, whereas IceCube has a footprint of a square kilometer. We performed calculations to determine the total number of telescopes and stations required to cover the full IceCube array over the 36$^\circ$ field of view. A hexagonal array of seven-telescope stations with a distance between stations of $\sim$260~m ensures that no point within the array is more than 150~m from the nearest station. Note that for air showers near this maximum distance from the nearest station, there will be several telescopes at comparable distances.  This provides several chances to detect the shower in an individual telescope, as well as the possibility of combining information from multiple stations to improve detection efficiency. To ensure coverage of the full field of view, the IACT array should extend several hundred meters beyond the edge of the IceCube footprint, as shown in Figure~\ref{fig:IceActLayout_fov18_white}.  

The full complement of seven telescopes is not required at each station.  For each telescope in each station, we calculated how many of IceCube's digital optiocal modules (DOMs) are located in the opposite direction of the telescope's FoV (i.e., how many DOMs the telescope would "protect").  The results are shown in Figure~\ref{fig:IceActLayout_fov18_white}, with the color of each telescope hexagon representing the number of DOMs behind the IACT.  A telescope array with a FoV of $36^\circ$ and a telescope station spacing of $\sim260$~m would include a total of 253 telescopes distributed among 83 stations.  This includes all telescopes that cover at least one DOM in their FoV. At an estimated cost of around 10,000~euros per telescope, this corresponds to an investment of order 3 million euros.  It would be possible to reduce the number of telescopes and stations required by eliminating telescopes protecting only the edges of the IceCube array, but detailed simulations of both the air showers and the IceCube detector response would be required to quantify the impact on veto efficiency.

The array shown in Figure~\ref{fig:IceActLayout_fov18_white} would cover a field of view of 0.27 sr (17$^\circ$ from zenith).  Based on estimates that at least several hundred sources contribute to the astrophysical neutrino flux observed by IceCube \citep{Ahlers:2014ioa}, this field of view should contain a number of as-yet-unidentified neutrino sources.  The field of view could be increased to 0.83 sr (30$^\circ$ from zenith, one fifteenth of the sky) by adding another ring to the fly's eye configuration at each station, increasing the number of potential telescopes at each station to 19.  For this scaled-up telescope array, a total of 723 telescopes distributed over 151 stations would be needed to cover the footprint of IceCube using the same assumptions above.  The estimated cost of this expanded array would still be less than 10 million euros.

\section{Impact on Neutrino Astronomy} \label{sec:neutrinoastronomy}

\begin{figure}[t!]
    \centering
  \resizebox{0.65\hsize}{!}{\includegraphics{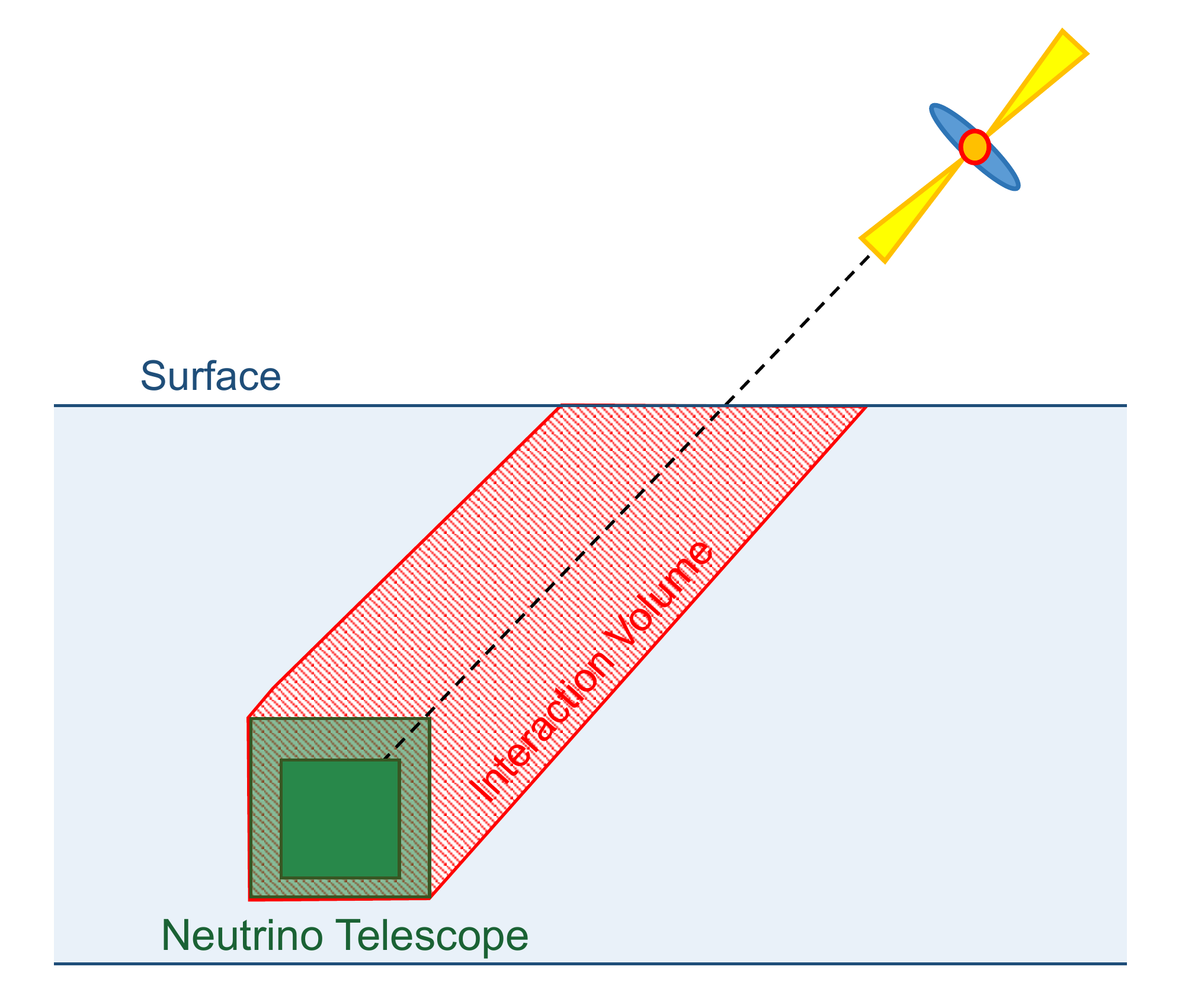}}
  \caption{Volumes for accepted neutrino interactions for a source at a given zenith angle.  For the standard starting event analysis, using the outer region of the detector as a veto, all neutrinos must interact in the fiducial volume indicated by the solid shaded region.  With a surface veto array, neutrinos interacting in the outer (shaded) region of the detector can also be accepted.  For long tracks from $\nu_\mu$ CC events, the allowed interaction volume could be extended to the ice above the detector as well.}
  \label{fig:trackinteractionvolume}
\end{figure}

The standard IceCube search for neutrinos from the overhead sky \citep{bib:HESE} employs a veto cut that excludes the outer volume of the detector in order to veto atmospheric muons.  This reduces the fiducial volume in which neutrinos are allowed to interact by approximately 40\%.  With a veto based on an IACT telescope array, neutrinos interacting anywhere inside the full volume of IceCube could be used, providing a full km$^3$ fiducial volume for cascades (charged current $\nu_{e,\tau}$ and neutral current $\nu_{e,\mu,\tau}$ events).  For charged current $\nu_\mu$ interactions, which produce long muon tracks, the usable interaction volume would include ice above the detector as well as the full IceCube detector volume, as shown in Figure~\ref{fig:trackinteractionvolume}.  This would increase the available interaction volume for tracks, and thus the event rate from an overhead neutrino source, by a factor of 4--5, depending on the zenith angle.

To illustrate the increase in sensitivity to neutrino sources obtained by adding an IACT telescope array to IceCube, we consider a source located at a zenith angle of 15$^\circ$,
and assume a flux similar to that emitted by the TXS 0506+056 blazar in 2015--16 \citep{bib:TXSBlazar}.  The average $\nu_\mu + \bar{\nu}_\mu$ flux observed from TXS 0506+056 over a 158 day period had a normalization at 100~TeV of $\Phi_{100}=1.6\times 10^{-18}$ $\mathrm{GeV}^{-1} \; \mathrm{cm}^{-2} \; \mathrm{s}^{-1}$ and a spectral index of 2.2. 

We estimate that a surface array such as that outlined in Section~\ref{sec:array} would detect air showers effectively above a cosmic-ray energy threshold of 50~TeV, corresponding to an atmospheric neutrino energy threshold of roughly 15~TeV.  For comparison, the neutrino energy threshold above which the atmospheric background is generally absent in the standard IceCube high-energy starting event analysis (using the periphery of the detector as a veto) is around 100~TeV \citep{bib:HESE}.  For the purposes of this study we assume that atmospheric neutrino backgrounds are reduced to negligible levels above this threshold; detailed simulations with high statistics will be required to quantify the exact atmospheric background rate surviving the veto.  Complete elimination of the atmospheric background is not necessary to provide substantially improved sensitivity to astrophysical neutrino sources.  A realistic analysis would likely weight individual events according to their likelihoods of being missed by the IACT veto, and would exploit neutrinos at energies below the threshold given here.  Our model of the veto efficiency as a step function at a precise energy threshold is intended only as a simplistic first estimate of the potential benefits of the array.

\begin{figure}[tp!]
     \centering
     \begin{subfigure}{\textwidth}
         \centering
         \includegraphics[width=0.65\textwidth,center]{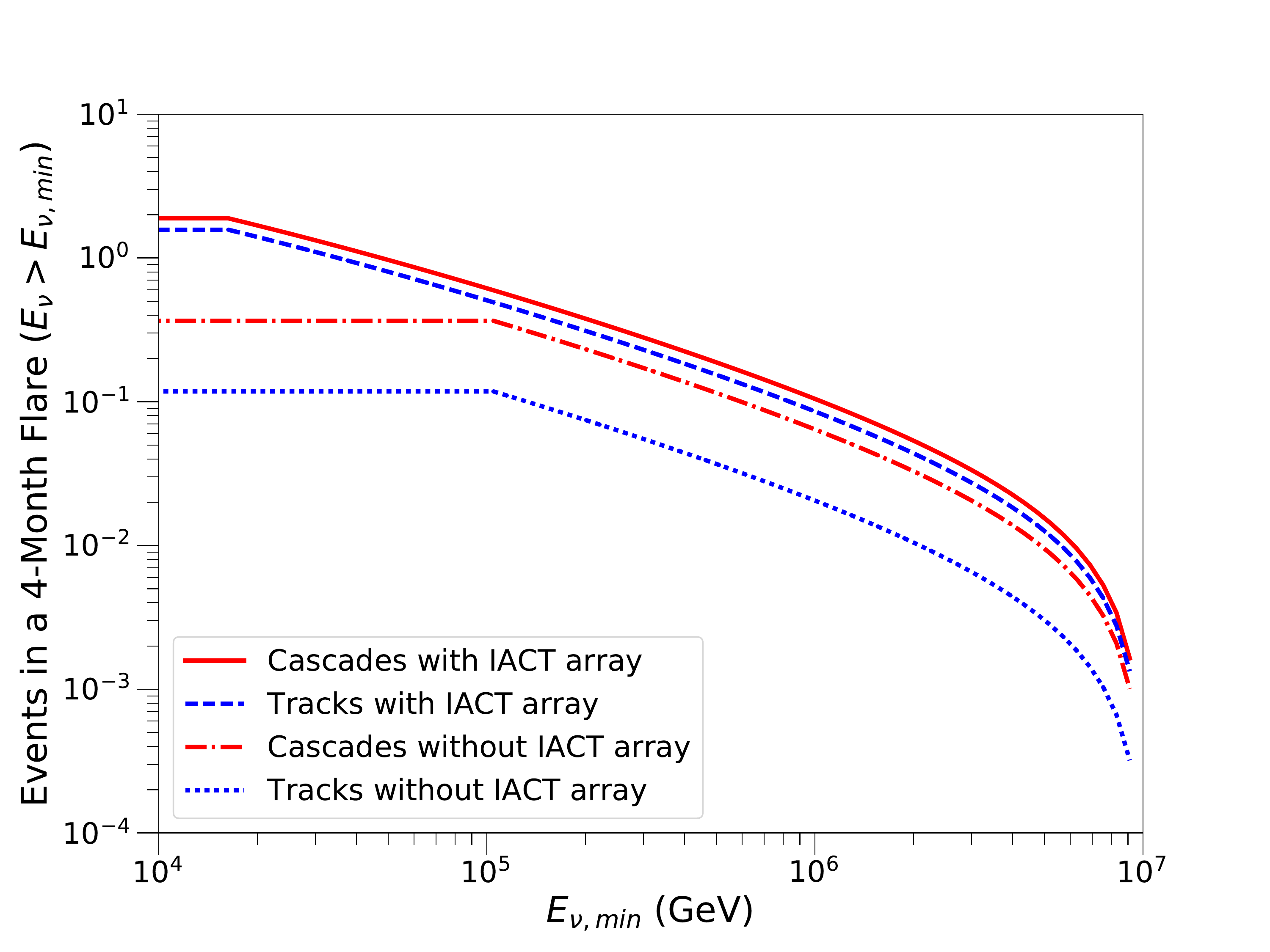}
         \caption{Number of identified neutrino events during a 4-month flare during the winter when the IACT surface array has a 60\% duty cycle.}
         \label{fig:4monthflare}
     \end{subfigure}
     \par\bigskip
     \begin{subfigure}{\textwidth}
         \centering
         \includegraphics[width=0.65\textwidth]{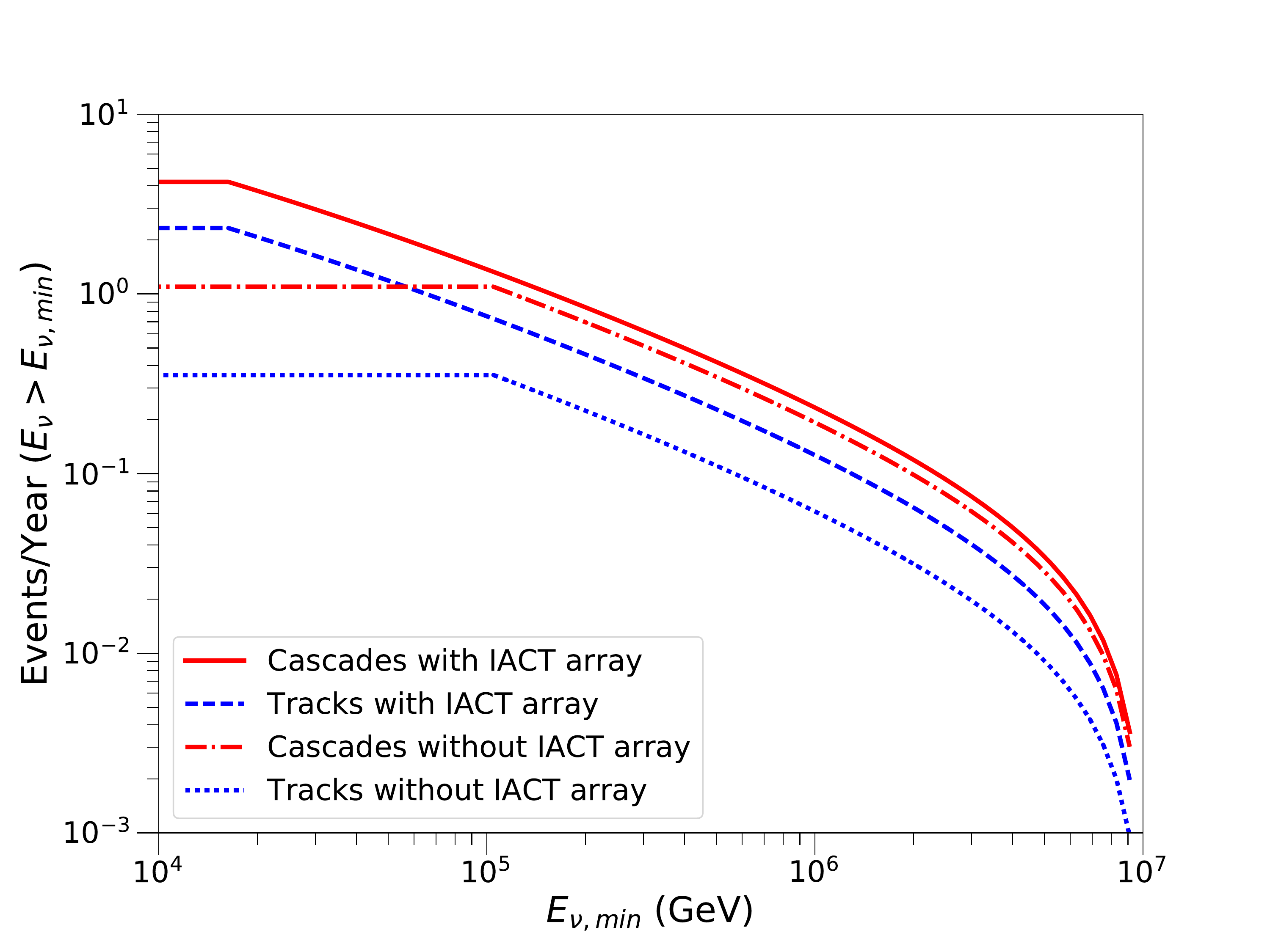}
         \caption{Number of identified neutrino events during a year-long flare when the IACT surface array has a 20\% duty cycle.}
         \label{fig:yearflare}
     \end{subfigure}
        \caption{Number of neutrino events passing the veto for a point source similar to TXS 0506+056 at a zenith of $15^{\circ}$. $E_{\nu,min}$ is the minimum neutrino energy emitted by the source.  When the detector veto energy threshold is above the minimum emitted energy, the number of events is independent of $E_{\nu,min}$. The IACT surface array primary energy threshold is taken to be 50~TeV, corresponding to a neutrino energy of 15~TeV, and the  neutrino energy threshold for the standard veto is 100~TeV.}
        \label{fig:neutrinorates}
\end{figure}

The numbers of astrophysical neutrinos observed above the veto threshold from a point source at a zenith of $15^{\circ}$ emitting a $\nu_\mu + \bar{\nu}_\mu$ flux of $1.6\times 10^{-7} \ E_{\nu}^{-2.2}$ $\mathrm{GeV}^{-1} \mathrm{cm}^{-2} \mathrm{s}^{-1}$ over two different durations are shown in Figure~\ref{fig:neutrinorates}.  Tracks refer to charged current $\nu_\mu$ events.  Cascades include charged current $\nu_{e,\tau}$ and neutral current $\nu_{e,\mu,\tau}$ events.  Because the source is in the Southern sky, neutrino interactions must take place within the 0.6~km$^3$ fiducial volume to pass the standard veto analysis; without the IACT veto array, tracks entering from outside the volume are always assumed to be atmospheric.  With the IACT veto, tracks produced by neutrinos interacting in the ice above the detector can also be accepted since the lack of an air shower indicates they are not of atmospheric origin.  \footnote{This assumes that atmospheric muons as well as atmospheric neutrinos can be vetoed effectively by the IACT array.  Atmospheric muons should be easier to veto than neutrinos, for several reasons.  Since over half of the muon energy will be deposited in the ice before reaching the detector, the minimum muon energy required to reach the neutrino telescope with at least 15~TeV (and thus the air shower energy threshold) is higher.  Atmospheric muons are produced in high-multiplicity bundles, which are distinguishable in neutrino telescopes from the single muons produced in $\nu_\mu$ CC interactions, so the neutrino telescope response can be incorporated to improve the veto efficiency.  However, since the atmospheric muon rate is considerably higher than that of atmospheric neutrinos, this assumption remains to be validated with a detailed, high-statistics simulation of IceCube operating in conjunction with the IACT array.  If sufficiently high atmospheric muon veto efficiency cannot be achived, the relative gains for track events will be similar to those for cascades.}


Figure~\ref{fig:4monthflare} shows the number of neutrinos observed from the source over a 4-month flare during the winter, when the telescopes have a duty cycle of 60\%. During this type of flare, around a factor of 13 more tracks and a factor of 5 more cascades would be observed with the IACT surface array than without it. Although only a few events of each flavor are expected, two factors should be borne in mind.  First, the actual event rates will be Poisson-distributed around the means shown in Figure~\ref{fig:neutrinorates}; higher numbers of events would be observed from a substantial fraction of actual flares.  Second, the discovery of TXS 0506+056 was initiated by a single neutrino event.  Two high-energy tracks from the same location in a relatively short period of time, with evidence from the IACT veto favoring non-atmospheric origin, would be enough to initiate multi-messenger follow-up observations and enable potential discoveries.  Figure~\ref{fig:yearflare} shows the number of neutrinos from a TXS 0506+056-like source during a year-long flare when the telescopes have an overall duty cycle of 20\%. During this type of flare, around a factor of 6.5 more tracks and a factor of 3.8 more cascades would be identified with the IACT surface array than without it.  

\section{Conclusions}

An array of small, $O(0.25~\rm{m}^2$), wide field of view Cherenkov telescopes would provide an efficient detector of air showers for the purpose of vetoing atmospheric neutrinos in a high-energy neutrino telescope.  The air Cherenkov technique is complementary to extensive air shower arrays (e.g. IceTop), which offer higher duty cycle, but have difficulty detecting lower energy ($E_{p} \lesssim 300$~TeV) air showers or those produced high in the atmosphere.  With a camera based on silicon photomultipliers (SiPMs), an array deployed above the IceCube Neutrino Observatory would be capable of taking data during the four months of astronomical night available at South Pole including periods with bright moon or aurora australis.  Filters selecting UV light would greatly reduce the impact of the aurora, enabling operation in most auroral conditions.  An overall duty cycle, accounting for periods of poor visibility, of 20\%--25\% (2.5 to 3 months per year) is expected to be possible.

First estimates of the telescope response to air showers indicate that with a telescope spacing of approximately 260 m, efficient air shower detection is expected above primary cosmic-ray energies of 50--100~TeV, corresponding to rejection of atmospheric neutrinos above energies of 15--30 TeV.  These thresholds could be tuned by changing the spacing between stations in the array.   More detailed simulations of the telescopes in conjunction with the buried neutrino detector are required to refine these estimates further.  The IceCube telescope could be covered with a field of view of 36$^\circ$ (i.e., zenith angles up to 17$^\circ$) by an array consisting of approximately 250 telescopes at 80 stations.  Approximately three times as many telescopes and twice as many stations would be required to extend the field of view to 30$^\circ$ from zenith.  

At a cost scale of several million euros, such an array would significantly enhance the atmospheric neutrino veto capabilities of IceCube at energies as low as tens of TeV.  This would greatly aid measurements of the lower end of the astrophysical neutrino spectrum, which have significant impact on understanding of the neutrino-gamma ray energy budget and the contributions of accelerators other than blazars to the neutrino sky.  In addition, reduction of the atmospheric neutrino background would enhance the sensitivity of IceCube to neutrino emission from astrophysical objects, particularly transient multi-messenger emission such as that observed from the blazar TXS 0506+056.

\section*{Acknowledgements}This material is based upon work supported by the National Science Foundation under Grant No.~1707842.  Any opinions, findings, and conclusions or recommendations expressed in this material are those of the author(s) and do not necessarily reflect the views of the National Science Foundation.  This work has also been supported by Michigan State University, RWTH Aachen University, the Verbundforschung of the German Ministry for Education and Research (BMBF) and by the Excellence Initiative of the German federal and state governments and the Helmholtz Alliance for Astroparticle Physics (HAP).

\bibliographystyle{elsarticle-num} 
\bibliography{references}

\end{document}